\documentclass{article}
\PassOptionsToPackage{numbers,sort&compress}{natbib}
\usepackage[preprint]{neurips_2026}
\usepackage[T1]{fontenc}
\usepackage[utf8]{inputenc}
\usepackage{times}
\usepackage{graphicx,booktabs,amsmath,amssymb}
\usepackage[colorlinks=true,linkcolor=blue,citecolor=blue,urlcolor=blue]{hyperref}
\newcommand{\kjmol}{\mathrm{kJ\,mol^{-1}}}
\newcommand{\angstrom}{\mathring{\mathrm A}}

\title{
GPU infrastructure for ML/MM with electrostatic embedding for enzyme chemistry
}

\author{Venkat Kapil \\
Department of Physics and Astronomy, University College London \\
Thomas Young Centre and London Centre for Nanotechnology \\
\texttt{v.kapil@ucl.ac.uk}}
\begin{document}
\maketitle

\begin{abstract}
We present a fully GPU-native framework for machine learning/molecular mechanics (ML/MM) enhanced sampling simulations with electrostatic embedding, demonstrated on an enzymatic reaction. 
The framework combines a \texttt{JAX} port of the pretrained \texttt{MACE-POLAR-1-M} model, an electrostatic embedding scheme compatible with the \texttt{MACE-POLAR} architecture, and integration with \texttt{JAX-MD} and \texttt{PLUMED-JAX} for end-to-end compiled enhanced sampling. 
The implementation is developed using \texttt{GPT-6.1-Sol} within \texttt{Codex}, with reasoning effort set to high. 
Using replica exchange umbrella sampling, we calculate the free energy profile of nucleophilic substitution in haloalkane dehalogenase in approximately 200 H100 GPU hours, treating the entire protein and reactants with \texttt{MACE-POLAR-1-M-JAX} and most of the solvent with molecular mechanics.
\end{abstract}

\section{Introduction}
\noindent Simulating enzyme reactions requires describing both changes in chemical bonding and interactions with the surrounding protein and solvent.
Quantum mechanics/molecular mechanics methods address this challenge by treating a selected region with electronic structure theory and its surroundings with a classical forcefield~\cite{senn2009,li_qmmm_gpu4pyscf_2025}.
Replacing electronic structure calculations with machine learning interatomic potentials (MLIPs) can substantially reduce computational costs, but conventional MLIPs lack explicit electronic degrees of freedom, making it challenging to capture the influence of the environment on the charge distribution of the reacting region.
Furthermore, biochemical reactions often involve rare events on long timescales, requiring both efficient simulation implementations and enhanced sampling methods to obtain converged free energies.

\noindent We note that the first challenge can be addressed by the \texttt{MACE-POLAR} architecture, which conditions atomic features on the local electrostatic potential~\cite{polar2026}, allowing atoms, in principle, to respond to the surrounding molecular mechanics (MM) environment.
The use of electrostatic-potential features to describe long-range
interactions builds on the long-distance equivariant (LODE) representation
introduced by Grisafi and Ceriotti~\cite{grisafi_longrange_2019}.
The second can be addressed using \texttt{JAX-MD}~\cite{schoenholz_jax_2020} for efficient molecular dynamics and \texttt{PLUMED-JAX}~\cite{plumedjax}, a port of selected \texttt{PLUMED} functionality~\cite{tribello_plumed_2014,bonomi_promoting_2019}, for enhanced sampling.
To combine these capabilities in an end-to-end compiled graphics processing unit (GPU) workflow, we develop \texttt{MACE-POLAR-1-M-JAX}, a \texttt{JAX}~\cite{bradbury_jax_2018} implementation of the pretrained \texttt{MACE-POLAR-1-M} model, couple it to fixed MM charges, and integrate it with \texttt{JAX-MD} and \texttt{PLUMED-JAX}.
The implementation is developed using \texttt{GPT-6.1-Sol} within \texttt{Codex}, with reasoning effort set to high.
We describe these components separately before demonstrating their application to nucleophilic substitution in haloalkane dehalogenase.

\begin{figure}[tbp]
\centering
\includegraphics[width=0.75\linewidth]{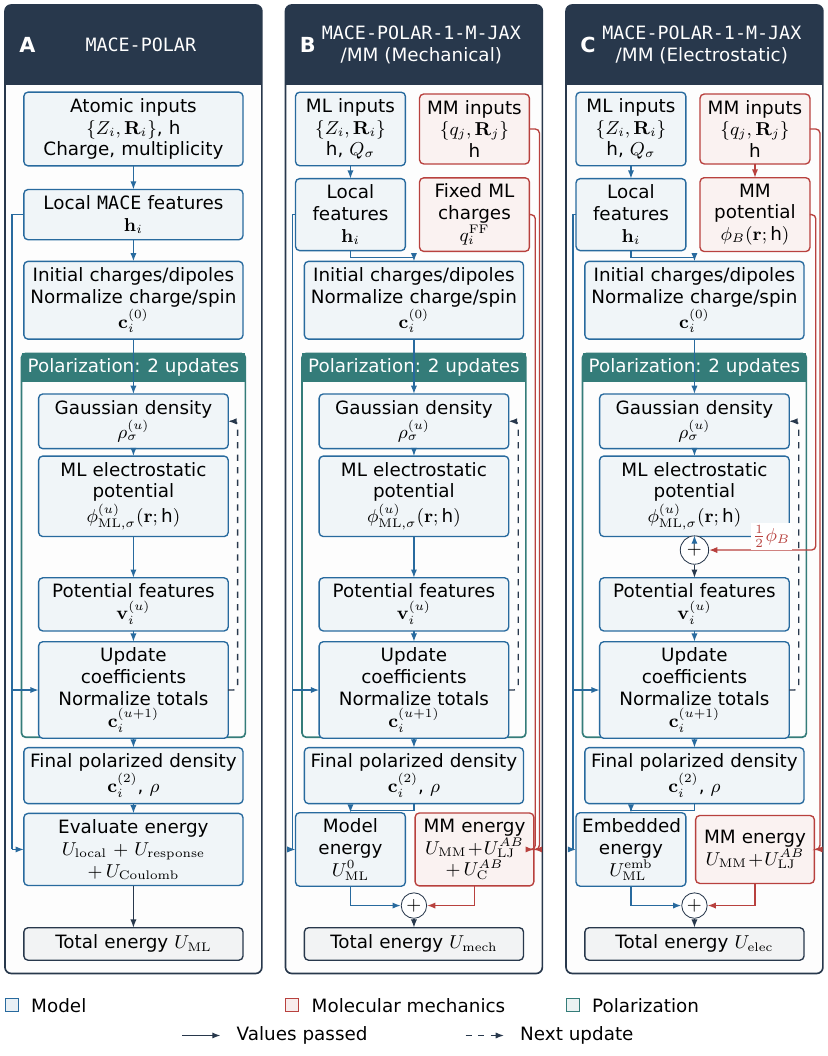}
\caption{
\texttt{MACE-POLAR} forward pass (A), mechanical embedding (B), and electrostatic
embedding (C).
Solid arrows show values passed and dashed arrows the next polarization update.
}
\label{fig:polar-schematic}
\end{figure}

\section{Methods and results}
\label{sec:methods-results}

\subsection{\texttt{MACE-POLAR-1-M-JAX}}

\noindent The \texttt{MACE-POLAR} architecture is summarized in
Figure~\ref{fig:polar-schematic}A.
The pretrained \texttt{MACE-POLAR-1-M} model~\cite{polar2026} instantiates this architecture with weights trained on Open Molecules 2025 data at the $\omega$B97M-V/def2-TZVPD
level~\cite{omol2025}.

\noindent Its inputs are the atomic numbers $Z_i$, positions $\mathbf R_i$, cell $\textsf{h}$, total charge, and spin multiplicity, where $i$ is the index of an atom.
Local \texttt{MACE} layers combine information from neighbouring atoms to describe their geometric environments~\cite{batatia_mace_2022}.
Learned functions of the resulting atomic features, denoted by $\mathbf h_i$ in the schematic, predict initial charges and dipoles in two spin channels.
Their coefficients are $\mathbf c_i^{(u)}$ at polarization update $u$, and
$\rho_\sigma$ denotes the density in spin channel $\sigma$.
Atom-centred Gaussian functions represent these coefficients as a smooth charge density, and a global correction enforces the specified total charge and spin multiplicity.
The model then calculates the electric potential generated by this density and integrates it against atom-centred Gaussian functions to obtain features describing the
potential around each atom.
These features include contributions from distant atoms and are combined with the local geometric features to update the charge and dipole coefficients.
The model performs two successive learned polarization updates, each using the potential generated by the current density and followed by a correction of the channel charges.
Finally, the energy combines local atomic contributions, a learned response contribution, and the electrostatic energy of the final density, including its interaction with an applied potential.
The mathematical formulation and pretrained model settings are given in the original
\texttt{MACE-POLAR} paper~\cite{polar2026}.

\noindent \texttt{MACE-POLAR-1-M-JAX} retains this forward pass and the pretrained weights. 
It supports tensor operations through the \texttt{cuEq} bindings for
\texttt{JAX}~\cite{nvidia_cuequivariance}.
Automatic differentiation supplies the forces and the derivatives used to calculate periodic stress.
Neighbour construction and model evaluation can be compiled together
using either the original reciprocal space Fourier sum or an additional
generalized particle-particle-particle-mesh (PPPM) solver.

\subsubsection{Numerical verification}

\noindent We compare \texttt{MACE-POLAR-1-M-JAX} with \texttt{MACE-POLAR-1-M} evaluated in \texttt{PyTorch}~\cite{paszke_pytorch_2019}, using the same checkpoint, input precision, neighbour graphs, and periodic images.

\noindent We validate the implementation using two sets of configurations:
\begin{enumerate}
    \item  For molecules, we use all 46 HR46 geometries~\cite{hickey2014} from the
GSCDB138 input archive~\cite{gscdb_archive}, centred on their molecular centroids with the supplied charges and spin multiplicities.
Each is evaluated at zero field and positive and negative fields along each
Cartesian axis, with magnitudes of $0.001$ and
$0.0005\,\mathrm{V\,\angstrom^{-1}}$, giving 13 fields and 598
molecule--field combinations. \\

\item For periodic liquid water, we use three frozen configurations at
each of six sizes: 192, 375, 648, 999, 1,536, and 3,000 atoms, with a volume
of $28.8209\,\angstrom^3$ per water molecule.
The 999-atom inputs comprise the 300~K water configuration released with the
\texttt{MACE-POLAR} study~\cite{polar2026} and two small rigid-molecule
perturbations of it.
For the other sizes, water geometries from that configuration are randomly
oriented around slightly displaced cubic-grid positions.
The periodic comparisons use the original reciprocal solver in both
programs, with a cutoff of approximately $1.85\,\angstrom^{-1}$.
\end{enumerate}

\noindent Table~\ref{tab:model-validation} reports differences between the
pretrained \texttt{MACE-POLAR-1-M} model evaluated using \texttt{PyTorch}
and \texttt{MACE-POLAR-1-M-JAX} using \texttt{cuEq}, separately in float64
and float32 floating-point arithmetic.
Energies are divided by the number of atoms before averaging over
configurations, while vector and tensor differences include every Cartesian
component.
The dipole comparison uses identical coordinates and origin conventions.
Stress is obtained from derivatives under a joint deformation of coordinates
and cell.

\begin{table}[!h]
\centering
\caption{
Numerical differences between \texttt{MACE-POLAR-1-M} in \texttt{PyTorch}
and \texttt{MACE-POLAR-1-M-JAX} using \texttt{cuEq}, for molecules containing
2--15 atoms and periodic water configurations containing 192--3,000 atoms.
RMS and Max denote root mean square and maximum absolute differences,
respectively; $e$ is the elementary charge.}
\label{tab:model-validation}
\small
\setlength{\tabcolsep}{1.4pt}
\begin{tabular}{lrrrr}
\toprule
& \multicolumn{2}{c}{\shortstack{\texttt{PyTorch}--\texttt{JAX}\\float64}}
& \multicolumn{2}{c}{\shortstack{\texttt{PyTorch}--\texttt{JAX}\\float32}} \\
\cmidrule(lr){2-3}\cmidrule(lr){4-5}
Quantity & RMS & Max & RMS & Max \\
\midrule
\multicolumn{5}{l}{\textbf{Molecules: 598 molecule--field combinations}} \\
Total energy (meV/atom)
& $5.77\times10^{-6}$ & $2.49\times10^{-5}$ & $0.200$ & $0.977$ \\
Forces (meV/\AA)
& $8.39\times10^{-6}$ & $5.07\times10^{-5}$ & $0.206$ & $2.46$ \\
Atomic charges ($e$)
& $1.73\times10^{-9}$ & $1.38\times10^{-8}$
& $6.58\times10^{-7}$ & $5.63\times10^{-6}$ \\
Model dipole ($e$\AA)
& $4.10\times10^{-9}$ & $1.77\times10^{-8}$
& $4.25\times10^{-6}$ & $2.93\times10^{-5}$ \\
\midrule
\multicolumn{5}{l}{\textbf{Periodic water: 18 configurations}} \\
Total energy (meV/atom)
& $1.14\times10^{-6}$ & $1.28\times10^{-6}$ & $4.25$ & $9.79$ \\
Forces (meV/\AA)
& $2.45\times10^{-5}$ & $4.54\times10^{-4}$
& $1.31\times10^{-3}$ & $0.0114$ \\
Atomic charges ($e$)
& $3.09\times10^{-10}$ & $1.01\times10^{-8}$
& $2.59\times10^{-8}$ & $6.56\times10^{-7}$ \\
Model dipole ($e$\AA)
& $3.82\times10^{-7}$ & $2.29\times10^{-6}$
& $1.02\times10^{-4}$ & $2.72\times10^{-4}$ \\
Stress (meV/\AA$^3$)
& $8.73\times10^{-8}$ & $2.41\times10^{-7}$
& $1.68\times10^{-5}$ & $6.15\times10^{-5}$ \\
\bottomrule
\end{tabular}
\end{table}

\noindent The float64 results show close numerical agreement across the
tested quantities.
The larger float32 differences in energies are dominated by rounding in the
large, position independent atomic reference, which does not affect forces.

\subsubsection{Performance efficiency in model evaluation and molecular dynamics throughput}

\begin{figure}[tbp]
\centering
\includegraphics[width=\linewidth]{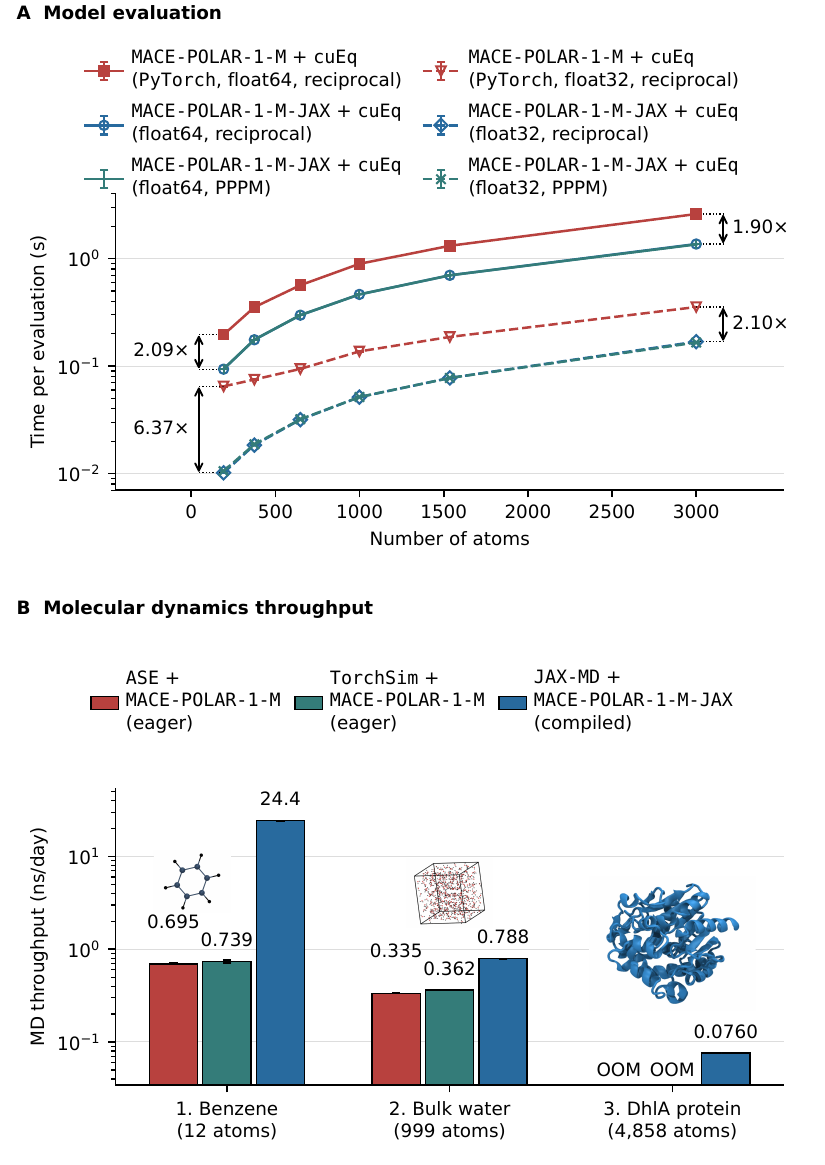}
\caption{
Computational performance on one RTX A6000: model evaluation (A) and
molecular dynamics throughput (B), with \texttt{cuEq} in both implementations.
Arrows in A compare the original reciprocal solver at matched precision;
error bars show the interquartile range of seven timing blocks.
Structures are rendered with \texttt{VMD}~\cite{humphrey_vmd_1996}, with
the water cell outlined and the protein shown as a backbone cartoon.
OOM denotes out of memory during the initial force evaluation.}
\label{fig:performance}
\end{figure}

\noindent We first measure the cost of a complete model evaluation,
including neighbour construction, graph assembly or transfer, energy,
forces, and stress, with outputs retained on the GPU.
Figure~\ref{fig:performance}A compares six periodic water sizes containing
192--3,000 atoms at $1.04\,\mathrm{g\,cm^{-3}}$ on one RTX A6000, using
\texttt{cuEq} and matched float64 or float32 arithmetic.
The original \texttt{MACECalculator} constructs neighbours on the central processing unit (CPU) and
transfers the graph to the GPU, whereas \texttt{JAX} compiles neighbour
construction and model evaluation together.
The tested \texttt{PyTorch} configuration runs in eager mode because it
does not support \texttt{torch.compile}.
Both implementations use the original reciprocal solver, with additional
\texttt{JAX} curves showing the cost of generalized PPPM.
Across the six sizes, \texttt{JAX} is 1.88--2.09 times faster in float64 and
2.10--6.37 times faster in float32, with respective speedups of 1.90 and
2.10 at 3,000 atoms, where PPPM and reciprocal timings are similar.

\noindent We then compare molecular dynamics throughput for isolated benzene
(12 atoms), periodic water (999 atoms), and isolated haloalkane dehalogenase
(DhlA; 4,858 atoms, charge $-17e$), where $e$ is the elementary charge
(Figure~\ref{fig:performance}B).
The pretrained model runs through \texttt{MACECalculator} with
the Atomic Simulation Environment
(\texttt{ASE})~\cite{larsen_ase_2017} and the official
\texttt{MaceTorchSimModel} interface with
\texttt{TorchSim}~\cite{cohen_torchsim_2025}, both in eager mode, while
\texttt{MACE-POLAR-1-M-JAX} uses \texttt{JAX-MD} with neighbour construction,
force evaluation, and integration compiled together.
All routes use \texttt{cuEq} on the same RTX A6000, float32 model
arithmetic, float64 integration, and a 0.5~fs velocity Verlet time step,
starting from identical positions and momenta initialized at 300~K,
without a thermostat, barostat, or constraints.
Water uses the original reciprocal solver, while benzene and the protein
use real space electrostatics; initial forces and trajectories are compared
for benzene and water to check the common settings.
Throughput is measured from seven blocks after warm-up, excluding loading,
compilation, trajectory writing, and thermodynamic logging.
Compiled \texttt{JAX-MD} reaches 24.4~ns/day for benzene and 0.788~ns/day
for water, respectively 33.0 and 2.17 times the \texttt{TorchSim} rates.
The larger gain for benzene is consistent with reducing host orchestration
around short force calculations.
The isolated protein runs at 0.0760~ns/day with
\texttt{MACE-POLAR-1-M-JAX}, while both tested \texttt{PyTorch}/\texttt{cuEq}
routes exhaust GPU memory during the initial force evaluation.

\subsection{Embedding \texttt{MACE-POLAR-1-M-JAX} in molecular mechanics}
\label{sec:embedding}

\subsubsection{Mechanical embedding}
\noindent A machine learning/molecular mechanics (ML/MM) calculation divides the atoms into a machine learning region $A$ and a classical molecular mechanics
region $B$, with positions $\mathbf R_A$ and $\mathbf R_B$.
The mechanical embedding approach evaluates region $A$ without the electric field of
region $B$ and adds classical interactions between the regions:
\begin{equation*}
U_{\mathrm{mech}}=
U_{\mathrm{ML}}^{0}(\mathbf R_A)+U_{\mathrm{MM}}(\mathbf R_B)
+U_{\mathrm{LJ}}^{AB}+U_{\mathrm{C}}^{AB}.
\end{equation*}
\noindent Here, $U_{\mathrm{ML}}^0$ is the model energy without the
classical environment, $U_{\mathrm{MM}}$ is the energy within region $B$,
and $U_{\mathrm{LJ}}^{AB}$ and $U_{\mathrm{C}}^{AB}$ are the cross-region
Lennard--Jones and electrostatic interactions.
The charges defining $U_{\mathrm{C}}^{AB}$ can be fixed forcefield
charges or geometry-dependent charges.
The latter allow charge redistribution along a reaction but not
polarization by the classical environment, as in the chorismate-mutase
calculations of Sun and co-workers~\cite{sun_mechanical_2026}.
Figure~\ref{fig:polar-schematic}B illustrates how mechanical embedding can be implemented with \texttt{MACE-POLAR}. 

\subsubsection{Electrostatic embedding}
\noindent Electrostatic embedding includes the response of the learned
region to its classical surroundings.
This response can be supplied by a separate physical model:
electrostatic machine learning embedding~\cite{zinovjev_embedding_2023}
combines a gas-phase potential with permanent electrostatics and coupled,
Thole-damped induced dipoles, as used by Gradisteanu and
co-workers~\cite{gradisteanu_enzyme_2026} and refined by Morado and
co-workers~\cite{morado_embedding_2025}.
Alternatively, environmental information can enter the MLIP:
Lei and co-workers~\cite{lei_continual_2025} use spatial derivatives of the classical potential, while \texttt{AMP-BMS/MM}~\cite{thurlemann_amp_2026} introduces field-induced dipole and quadrupole features during message passing, which is related to the approach used here.

\noindent Figure~\ref{fig:polar-schematic}C illustrates electrostatic embedding
with \texttt{MACE-POLAR}.
The potential $\phi_B$
generated by the fixed classical charges enters the learned polarization updates of \texttt{MACE-POLAR-1-M-JAX}.
It is shared equally between the two spin channels, whose total charges are denoted by $Q_\sigma$.
This allows the model charges and dipoles to respond to the environment even when the geometry of region $A$ is unchanged, while the classical charges remain fixed.
The total energy is
\begin{equation*}
U_{\mathrm{elec}}=
U_{\mathrm{ML}}^{\mathrm{emb}}(\mathbf R_A;\phi_B)
+U_{\mathrm{MM}}(\mathbf R_B)+U_{\mathrm{LJ}}^{AB}.
\end{equation*}
\noindent Here, $U_{\mathrm{ML}}^{\mathrm{emb}}$ includes the model energy,
its response to $\phi_B$, and the cross-region electrostatic interaction,
so no separate $U_{\mathrm{C}}^{AB}$ term is added.
In our implementation, \texttt{MACE-POLAR-1-M-JAX} supplies the embedded model energy, while \texttt{JAX-MD} evaluates the classical interactions.
Cross-region Lennard--Jones interactions use arithmetic mixing of size parameters and geometric mixing of well depths.
For periodic systems, generalized PPPM evaluates the Gaussian multipole potential and coupled electrostatic energy, while the classical region uses point-charge particle mesh Ewald~\cite{essmann1995}.

\subsection{Enzyme free energy calculations}
\label{sec:sampling}

\subsubsection{System}

\noindent To demonstrate the proposed ML/MM scheme, we consider the first substitution step catalysed by DhlA with 1,2-dichloroethane.
The carboxylate group of aspartate residue 124 (Asp124) attacks a substrate carbon while chloride departs, forming a covalent ester intermediate~\cite{plotnikov2014,schanstra1996}.
This process combines bond formation, bond breaking, and a changing charge distribution in a protein environment.

\noindent Plotnikov~\cite{plotnikov2014} calculated a B3LYP/MM free energy profile, and Schanstra and co-workers~\cite{schanstra1996} measured the rate of the chemical step, providing computational and experimental comparisons.
Following Ref.~\citenum{plotnikov2014}, we describe the progress of the reaction using $s=d(\mathrm{C{-}Cl})-d(\mathrm{C{-}O})$, where C is the attacked substrate carbon, Cl is the departing chlorine, O is the attacking carboxylate
oxygen OD2 of Asp124, and $d$ denotes an interatomic distance.
Negative values describe reactants, and positive values describe the ester intermediate, which is the product considered here.

\begin{figure}[tbp]
\centering
\includegraphics[width=\linewidth]{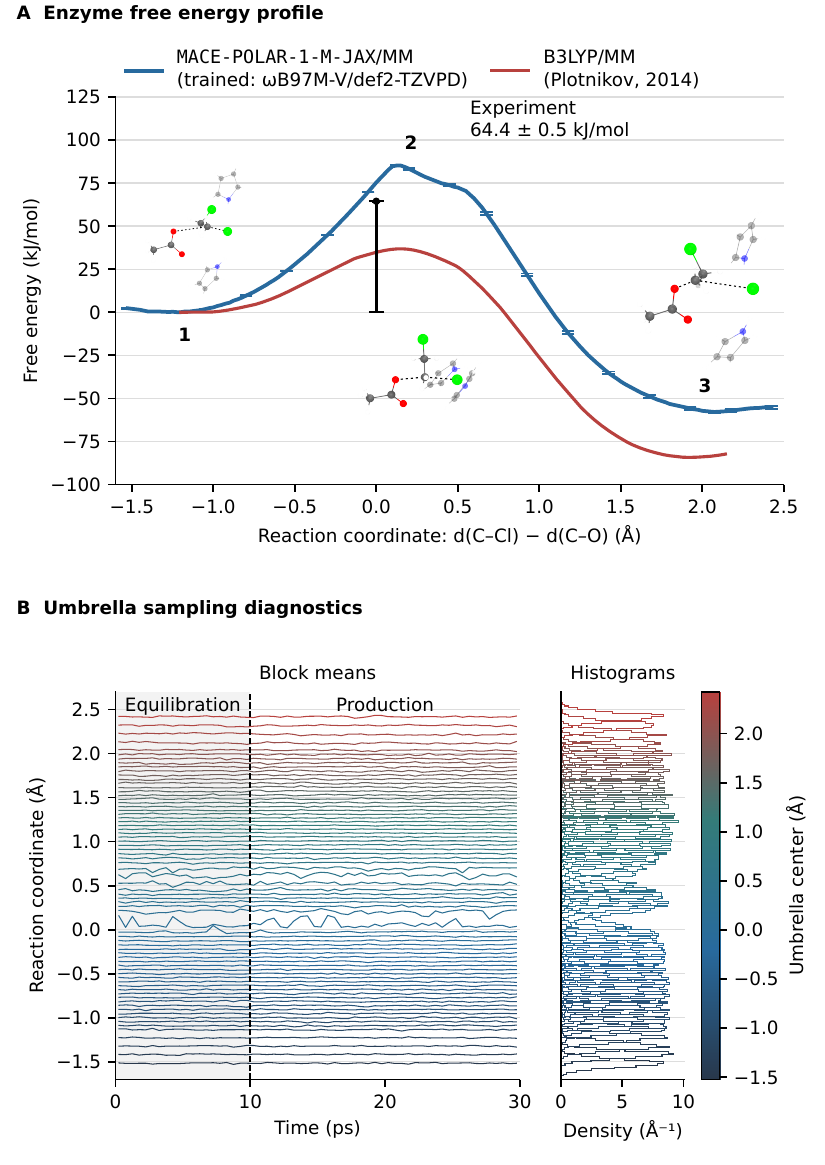}
\caption{
DhlA free energy profiles (A) and sampling diagnostics (B).
Blue in A shows electrostatic machine learning/molecular mechanics and red
the B3LYP/MM profile from Plotnikov~\cite{plotnikov2014}; shading and error
bars are block-jackknife standard errors, and the experimental marker is
rate-derived~\cite{schanstra1996}.
Snapshots rendered with \texttt{VMD}~\cite{humphrey_vmd_1996} show capped
reacting fragments and neighbouring residues at reactant (1), near-barrier
(2), and ester intermediate (3) geometries; snapshot 2 is not an optimized
transition state.
Panel B shows 0.5~ps block means and normalized production histograms for
all 72 umbrellas; the dashed line marks the end of 10~ps equilibration.}
\label{fig:enzyme-profile}
\end{figure}

\subsubsection{Molecular dynamics setup}
\noindent The solvated system is prepared from Protein Data Bank entry 2DHC and contains
33,305 atoms in a periodic cubic cell of side 70.8685~\AA.
The electrostatically embedded region contains the entire protein, the substrate, and two waters, giving 4,872 atoms with charge $-17e$ and spin multiplicity one.
The remaining 28,433 atoms are classical water and counterions, and the complete cell is neutral.

\noindent The classical solvent uses \texttt{TIP3P} parameters~\cite{jorgensen1983}, and nonbonded interactions use a 10~\AA\ cutoff and a $108^3$ electrostatic mesh.
The Gaussian multipole mesh has target spacing 0.5~\AA\ and assignment order eight.
Separate neighbour lists for modelled, classical, and cross-region pairs use a 0.5~\AA\ buffer and are updated when any atom in the local batch moves more than 0.25~\AA\ from its reference position.
The smooth distance-ordering restraint is based on the flexible inner region ensemble separator~\cite{rowley2012}, using distances from the Asp124 CG atom to the
water oxygens.
It acts when a smooth estimate of the largest distance among modelled waters exceeds a smooth estimate of the smallest distance among classical waters, with force constant
$5\,\kjmol\,\angstrom^{-2}$.
Classical waters are rigid under the SETTLE position and velocity constraints~\cite{miyamoto1992}, while modelled waters are flexible.

\subsubsection{Enhanced sampling setup}

\noindent Following Ref.~\citenum{plotnikov2014}, we use umbrella sampling to estimate the reaction free energy profile.  
However, we also use replica exchange, which attempts Metropolis exchanges between neighbouring umbrellas at a common temperature~\cite{sugita_multidimensional_2000}.
Using a setup similar to that of Ref.~\citenum{plotnikov2014}, we run 72 umbrellas for 10~ps equilibration and 20~ps production each, using Langevin dynamics with a target temperature of 300~K.
The 64 central umbrellas have centres from $-1.125$ to $2.025$~\AA\ in increments of 0.05~\AA.
Four additional umbrellas at each end cover $-1.525$ to $-1.225$~\AA\ on the reactant side and $2.125$ to $2.425$~\AA\ on the product side,
both in increments of 0.10~\AA.
All 72 umbrellas use a force constant of
$1046\,\kjmol\,\angstrom^{-2}$.
Dynamics use a 0.5~fs time step and Langevin friction $1\,\mathrm{ps}^{-1}$.
Exchanges are attempted every 200 steps, and the reaction coordinate is recorded every 20 steps.
The free energy surface is estimated using the multistate Bennett acceptance ratio approach~\cite{shirts_statistically_2008}.

\subsubsection{Distributed execution}
\noindent We execute all 72 umbrellas in a single job, using
\texttt{JAX} sharding across 18 NVIDIA H100 GPUs, with four umbrellas
batched on each GPU using \texttt{jax.vmap}.
Within each umbrella, \texttt{JAX-MD} advances the dynamics with biases
supplied by \texttt{PLUMED-JAX}, using \texttt{jax.lax.scan} inside blocks
compiled by \texttt{jax.jit}.
Replica exchanges swap umbrella labels rather than transferring atomic
configurations between devices, with bias forces updated after each
accepted exchange.
\noindent Sampling all 72 umbrellas requires approximately 200~H100 GPU hours, including initialization, compilation, 10~ps equilibration, and 20~ps production per umbrella.

\subsubsection{Free energy profile and comparisons}

\noindent Figure~\ref{fig:enzyme-profile}A shows the reconstructed free energy profile relative to the reactant minimum.
The forward barrier is $85.2\pm0.4\,\kjmol$, and the ester intermediate minimum lies $57.9\pm0.9\,\kjmol$ below the reactant minimum.
The B3LYP/MM profile digitized from Figure~8 of
Plotnikov~\cite{plotnikov2014} gives a barrier of $36.8\,\kjmol$ and an approximate product--reactant difference of $-84.1\,\kjmol$, without reported statistical uncertainties.
The experimental chemical-step rate of $50\pm10\,\mathrm{s}^{-1}$ at
303.15~K~\cite{schanstra1996} corresponds to a barrier of
$64.4\pm0.5\,\kjmol$ using transition state theory with transmission coefficient one and uncertainty propagated from the rate.
Our barrier exceeds this rate-derived estimate by $20.8\,\kjmol$.
For comparison, Gao and co-workers~\cite{gao_enzyme_2026} report discrepancies ranging from $3.84\,\kjmol$ for poly(ethylene terephthalate) hydrolase to $17.5\,\kjmol$ for nucleoside diphosphate kinase, the latter approaching the discrepancy observed here
(Table~\ref{tab:experimental-context}).
Their calculations use the pretrained \texttt{eSEN-omol} potential, trained at the same $\omega$B97M-V/def2-TZVPD reference level, to describe both enzyme and solvent.
They also use the Eyring relation with transmission coefficient one to relate rates and barriers, although the experimental estimate for nucleoside diphosphate kinase comes from a turnover rate and represents an upper bound on its chemical barrier.

\begin{table}[!h]
\centering
\caption{
Calculated and rate-derived experimental barriers in kJ/mol, with calculated
minus experimental differences obtained from unrounded values.
DhlA uncertainties are a block-jackknife standard error and propagated
rate uncertainty.
Other uncertainties are reproduced from Gao et al.~\cite{gao_enzyme_2026};
for poly(ethylene terephthalate) hydrolase, the calculated uncertainty is the
half-width of the reported 95\% interval.
The nucleoside diphosphate kinase experimental estimate comes from a
turnover rate and is an upper bound on the chemical barrier.}
\label{tab:experimental-context}
\small
\begin{tabular}{lrrr}
\toprule
System & Calculated & Rate-derived & Difference \\
\midrule
DhlA, this work
& $85.2\pm0.4$ & $64.4\pm0.5$ & $20.8$ \\
Chorismate mutase
& $75.2\pm2.72$ & $64.4$ & $10.8$ \\
\shortstack[l]{Poly(ethylene terephthalate) hydrolase}
& $79.0\pm4.20$ & $75.2\pm0.4$ & $3.84$ \\
\shortstack[l]{Nucleoside diphosphate kinase (umbrella sampling)}
& $73.1\pm1.86$ & $55.6$ & $17.5$ \\
\bottomrule
\end{tabular}
\end{table}

\noindent Beyond the barrier height, the shape of our free energy
profile differs from the reference, with a sharp change in slope near
the maximum (Figure~\ref{fig:enzyme-profile}A).
To examine sampling near this feature,
Figure~\ref{fig:enzyme-profile}B shows the evolution of the reaction
coordinate and its production distribution in each umbrella.
The change in slope coincides with reduced overlap between neighbouring
umbrellas: the overlap between those centred at 0.125 and 0.175~\AA\
is 0.34, compared with a median of 0.62 across neighbouring pairs.
Here, overlap is calculated as
$O_{ij}=\sum_b\min(p_{ib},p_{jb})$, where $p_{ib}$ is the fraction of
production samples from umbrella $i$ in a common bin of width 0.02~\AA.
This correlation suggests that insufficient overlap could introduce a
systematic error in the reconstructed profile, although testing this
possibility requires another set of umbrella sampling simulations.

\section{Conclusions}
\noindent We have presented GPU infrastructure for electrostatic
machine learning/molecular mechanics simulations of enzyme reactions.
This combines a \texttt{JAX} port of the pretrained
\texttt{MACE-POLAR-1-M} model, an embedding scheme that allows the
classical environment to polarize the learned region, and integration
with \texttt{JAX-MD} and \texttt{PLUMED-JAX} for compiled enhanced
sampling.
Numerical comparisons quantify agreement with the original model,
while timings demonstrate faster model evaluation and molecular
dynamics for the tested systems.

\noindent Applied to nucleophilic substitution in DhlA, this framework enables
replica exchange umbrella sampling with the entire protein and reactants
in the learned region and most solvent described by molecular mechanics,
at a total cost of approximately 200 H100 GPU hours.
While the free energy estimates presented here are preliminary, their
deviation from experiment is comparable in scale to discrepancies
reported in other enzyme simulations using pretrained machine learning
potentials~\cite{gao_enzyme_2026,thurlemann_amp_2026}.
This computationally efficient infrastructure opens opportunities for systematic sampling
and comparisons of how protein conformations and environmental
polarization influence enzyme reactions.

\section*{Data and software availability}
\noindent The data supporting this report will be made available at a later stage.

\section*{Acknowledgments}
\noindent The author thanks members of the Initiative for Computational
Catalysis and the Center for Computational Biology at the Flatiron
Institute for insightful discussions.
Computational resources were provided by the Flatiron Institute, a division of the Simons Foundation.

\end{document}